\def\year{2022}\relax
\pdfoutput=1
\documentclass[letterpaper]{article} % DO NOT CHANGE THIS
\usepackage{aaai22}  % DO NOT CHANGE THIS
\usepackage{times}  % DO NOT CHANGE THIS
\usepackage{helvet}  % DO NOT CHANGE THIS
\usepackage{courier}  % DO NOT CHANGE THIS
\usepackage[hyphens]{url}  % DO NOT CHANGE THIS
\usepackage{graphicx} % DO NOT CHANGE THIS
\usepackage{natbib}  % DO NOT CHANGE THIS AND DO NOT ADD ANY OPTIONS TO IT
\usepackage{caption} % DO NOT CHANGE THIS AND DO NOT ADD ANY OPTIONS TO IT
\DeclareCaptionStyle{ruled}{labelfont=normalfont,labelsep=colon,strut=off} % DO NOT CHANGE THIS
\usepackage{algorithm}
\usepackage{algorithmic}

\newtheorem{remark}{Remark}
\usepackage{newfloat}
\usepackage{listings}
\floatstyle{ruled}
\newfloat{listing}{tb}{lst}{}
\floatname{listing}{Listing}
\nocopyright
\title{Market Design for Drone Traffic Management}
\author{
    Sven Seuken,\textsuperscript{\rm 1,\rm 2}
    Paul Friedrich,\textsuperscript{\rm 1,\rm 2}
    Ludwig Dierks\textsuperscript{\rm 1,\rm 2,\rm 3}
}
\affiliations{
    \textsuperscript{\rm 1}University of Zurich\\
    \textsuperscript{\rm 2}ETH AI Center\\
    \textsuperscript{\rm 3}Kyushu University
    \{seuken,paul,dierks\}@ifi.uzh.ch
}

\begin{document}

\maketitle

\begin{abstract}
The rapid development of drone technology is leading to more and more use cases being proposed. In response, regulators are drawing up drone traffic management frameworks. However, to design solutions that are \textit{efficient}, \textit{fair}, \textit{simple}, \textit{non-manipulable}, and \textit{scalable}, we need \textit{market design} and \textit{AI} expertise. To this end, we introduce the drone traffic management problem as a new research challenge to the market design and AI communities. We present five design desiderata that we have derived from our interviews with stakeholders from the regulatory side as well as from public and private enterprises. Finally, we provide an overview of the solution space to point out possible directions for future research.
% Market design and multi agent systems expertise is needed to develop UTM solutions that are fair, efficient, incentive compatible and scalable. To gain an understanding of the UTM environment, we have interviewed stakeholders from the regulatory side as well as from public and private UAV enterprises. This paper poses a research question to the AI and market design community by providing an analysis of the complexities of UTM system design, presenting market design desiderata specific to it and exploring the solution space.
\end{abstract}

\section{Introduction} 

\textit{Air traffic management (ATM)} is currently confronted with the rise of \textit{unmanned aerial vehicles (UAVs)}, so-called \textit{drones}. Advances in battery and materials technology have led to smaller and more agile drones that can carry cargo and equipment \cite{floreano2015science}. Better communications technology and advances in AI have led to higher drone autonomy and lowered the barrier to entry. This has led to an increase in the number of private drone pilots and spurred R\&D efforts into commercial applications (e.g., drone-based delivery services) \cite{HBR1, McKdrones2}.

%In the coming years, a further radical increase in both the number of UAVs \cite{doole2020estimation} and the breadth of use cases \cite{TODO: some analyst pieces (harvard etc)} is .
%This has led to an increase in the number of private drone pilots and spurred R\&D efforts (e.g., on drone-based delivery services), with the number of UAVs expected to radically increase in the coming years \cite{doole2020estimation}. 

Although the airspace for drones is not congested yet, the number of UAVs is expected to radically increase in the coming years \cite{doole2020estimation}. It is only a matter of time until drone operators will be competing for airspace, at least in certain locations (e.g., at desirable landing and take-off spots). Scaling up UAV usage to such levels requires well-designed \textit{UAV traffic management (UTM) systems} (i.e., mechanisms that allocate airspace to UAVs).\footnote{In Europe, UTM is commonly referred to as \textit{U-space}.} However, such systems are still in their infancy \cite{barrado2020u, thipphavong2018urban}. Designing suitable mechanisms is challenging, given that drones have diverse abilities and drone operators may have diverse preferences. To date, no solution has emerged that convincingly trades off system design desiderata like \textit{efficiency}, \textit{fairness}, \emph{simplicity}, \emph{incentives}, and \emph{scalability}.

\insert\footins{\noindent\footnotesize \\ \\}

While governments around the world are currently revising their UTM regulations (e.g., \citet{ECUspace, FAAconops, FOCAconops}), most of these draw heavily on existing regulations for \textit{manned} ATM, neglecting the particularities of the UAV space. Most importantly, the issue of \emph{strategic behavior} among operators has been largely ignored. This is understandable, given that in (traditional) manned ATM, there are only a few, comparatively similar companies (i.e., airlines), and uncooperative behavior can easily be reprimanded. However, there will be thousands of drone operators, and many of them can be expected to be \emph{self-interested} (i.e., aiming to maximize their own utility). Most UTM proposals allow operators to ``reserve'' parts of the airspace for free, with the option of cancelling their reservation. The most recent EU regulations, for example, foresee a \emph{first-come-first-serve} mechanism for these reservations \cite{ECUspace}. Unfortunately, once the airspace is getting congested, such a mechanism would mean that any strategic operator should hurry to reserve as much airspace as they can, as early as possible.\footnote{We further discuss shortcomings of the first-come-first-serve mechanism in Section \ref{md-desiderata}.} Of course, one could limit how many reservations an operator can make, or how often they can freely cancel. However, such ``fixes'' often introduce their own issues and should not be applied naively. Instead, any UTM proposal should be carefully analyzed before it is employed in practice.  % it is not easily apparent how to apply them without further endanger the efficiency or fairness of the UTM system.

%Thus, once the airspace is getting congested, a strategic operator should hurry to reserve as much airspace as they can, as early as possible. 

%Of course, one could limit how many reservations an operator can make, or how often they can freely cancel. But such ``fixes'' can easily endanger the efficiency or fairness of the UTM system.

% regulatory bodies internally often lack the expertise to evaluate the strategic implications of newly proposed rules. They are further heavily influenced by manned aviation experience and priorities, presenting a  a clear danger of drawing too heavily from comparatively monolithic manned ATM and neglecting particularities of the more dynamic and diverse UAV space. One often proposed rule, allocating airspace on a first-come-first-serve basis, for instance, distorts the market by incentivizing agents to book substantial parts of the airspace in advance, leaving little room for urgent flights with short planning horizons. 

%This poses an opportunity for the market design and multi-agent systems community to fill these knowledge gaps and help improve UTM regulations around the globe.

%%%%%%%%%%%%%
% - mention multi-agent systems
% - mention AI
% - mention game theory
% - mention mechanism design
% - mention market design
% - mention optimization
% - mention planning/scheduling
% - mention behavioral considerations

We argue that techniques from \emph{market design} as well as \emph{AI} are most suited to design a UTM system that satisfies all important desiderata (like efficiency and fairness) while also being scalable in practice. Market design is a research area at the intersection of computer science, economics, and operations research, that focuses on the design of well-functioning marketplaces \citep{kominers2017MarketDesign}.\footnote{The term \textit{market} or \emph{marketplace} refers to a system of rules that allocates goods to agents based on their actions. It does not need to (but can) contain monetary payments (see Section \ref{marketbasedsolutions}).} In particular, using tools from game theory and mechanism design \citep{myerson2008}, it explicitly takes the strategic behavior of agents into account and aims to design rules that nonetheless lead to good outcomes. Researchers from AI (in particular, from multi-agent systems) have experience with tackling large-scale coordination tasks that require solving complex optimization problems \citep{wooldridge2009MultiAgentSystems}. Over the last 25 years, researchers from market design and AI have successfully tackled some of the most complex resource allocation problems. Prominent examples include spectrum auctions \cite{cramton2013SpectrumAuctionDesign}, the allocation of courses to students at universities \cite{budish2017course}, and the allocation of vaccines to countries \cite{camilo2021MarketDesignCovid}. We believe that the design of a good UTM system is a similarly important and exciting challenge.

To understand the challenges in UTM design and derive the most important desiderata, we have interviewed ten stakeholders from regulatory bodies as well as public and private enterprises (see the \textit{Acknowledgements} at the end of this paper).
%In this paper, we introduce the UTM design problem to the research community.
We first describe the UTM design problem in Section \ref{problemoverview}, before discussing the most relevant market design desiderata in Section \ref{md-desiderata}. Finally, in Section \ref{solutionspace}, we sketch the space of potential solutions.   

\section{Problem Overview} \label{problemoverview}

The \textit{airspace} is a 4-D space, consisting of three spatial dimensions and one time dimension. \textit{UAV operators} arrive over time and want to complete flights, and each operator has preferences for when and where they fly. A flight is a \mbox{4-D} trajectory through the airspace. A UTM \emph{mechanism} allocates airspace to self-interested operators.
A \emph{feasible} allocation ensures that no two flight paths intersect. Each operator has a certain \textit{value} for their optimal path, but most operators can accommodate deviations at a \textit{cost}. Operators' values and costs are their private information (i.e., only known to themselves) and they act strategically in reporting them. 

Drone operators greatly differ in the goals and flexibility of their flights. For example, a wedding photographer may place a high value on flying their drone directly above the wedding, at the precise time the newlyweds exit the building, while any deviation in space or time would drastically reduce their value. In contrast, a delivery drone operator may not care about the exact trajectory of a flight, as long as they reach their destination within a certain time window.

Operators also differ in how they can plan and execute flights, making the \textit{timing} of allocations particularly important. Some operators may want to receive their allocations far in advance (e.g., to plan a photo shoot), while others only know their needs shortly before take-off (e.g., food delivery services) or may desire nondisclosure until the last minute. Operators may have preferences that span multiple flights, e.g., wanting to photograph a certain area on one of several consecutive flights. Preferences may shift as time goes on, and operators may want to submit new flight plans at any point in time. Additionally, while current proposals focus on strategic (pre-flight) allocation and try to keep tactical (in-flight) re-allocation of airspace to a minimum, a UTM mechanism still needs to be robust against events that make dynamic re-allocation unavoidable (e.g., if battery failure or a thunderstorm make flying a particular route impossible).

Any UTM mechanism must also take multiple \emph{practical considerations} into account \cite{lundberg2018urban}. For example, stakeholders agree that public-interest flights (e.g., emergency services, police) must always be \textit{prioritized} over regular participants. Additionally, to ensure \emph{safety}, UAV flight paths need to maintain a minimum spatial and temporal separation from each other and from potential exogenous obstacles (e.g., buildings or bird swarms). %Not all constraints are static, and a UTM mechanism should be robust against events that suddenly limit airspace access (e.g., weather) or make flights infeasible (e.g., battery failure). 
This is complicated by the heterogeneous ability of drones and operators to detect and react to threats or changes in their flight plan and varying degrees of aerial maneuverability (e.g., not every drone can ``hover'' or turn in place). Furthermore, drones differ greatly in their operational range (e.g., some drones may need to maintain line of sight) and maximum flight times (e.g., due to battery capacity). Contrary to the name, not all UTM participants may actually be ``unmanned'' and any future-proof UTM design needs to also take applications such as (autonomous) air taxis into account. To ensure that proposed UTM systems can cope with any future developments, the market design should be as technology agnostic as possible. However, to operate the market in practice, it is important that UTM designers still take technological constraints into account when specifying regulations. Considering the heterogeneity of the domain, it is desirable that the system is able to directly query individual operators' constraints and requirements. 

Depending on the jurisdiction, there may also be \textit{politically imposed side-constraints} regarding noise pollution, privacy, and safety. Any drone flight, no matter how strict the safety regulations, leads to certain risks; concretely, a flight poses a small danger to other drones, to people, and to the infrastructure on the ground. Of course, drone operators have some inherent incentive to fly safely to minimize these risk, because an accident would likely be costly and cumbersome for them and could also lead to stricter regulations being imposed on the drone market as a whole. However, this has limits, as some companies might be more risk-seeking or have shorter planning horizons than their competitors or the regulator. Furthermore, accidents that lead to large collateral damage might simply bankrupt smaller players. Consequently, these players may not fully internalize the externalities they create, forcing a socialization of the resulting costs. Thus, it is important that the regulator defines a \textit{minimum level of safety} on which they are not willing to compromise and which therefore constitutes a hard market design \textit{constraint}.  One way to think about this constraint is that safety requirements below this level would endanger the overall health of the drone market and, in the worst case, lead to complete market failure. While higher levels of safety above this minimum are still in the regulator's (and most operators') interest, at that point safety turns from a simple constraint into a desideratum (where more is better) that must be carefully traded off against the other market design desiderata we introduce in the following section.

\section{Market Design Desiderata} \label{md-desiderata}
%In this section, we present five desiderate that 
Based on our stakeholder interviews, we have identified five desiderata that are essential for any UTM mechanism. Not all of these desiderata can be maximized simultaneously (e.g., \citet[Chapter 2]{parkes2001PhDThesis}), such that any UTM design must carefully trade them off against each other.

%Safety can be seen as both a constraint and desideratum. It is certainly a core requirement of any UTM system, but one which presents tradeoffs. Increasing safety may lower the efficiency of a system (e.g., disallowing routes over pedestrian areas making average routes longer). It could also impact fairness, for instance, if safety measures reduced all agents’ utilities, thus leading to a lower egalitarian social welfare.

\begin{remark}
There are three different notions for most of these desiderata: \emph{ex-ante}, \emph{ex-interim}, or \emph{ex-post} \cite{mas1995microeconomic}. Ex-ante means that a desideratum holds \textit{in expectation}, given some probability distribution of future world states. Ex-interim means that a desideratum holds while parts of the world have already been realized (e.g., the preferences of a single agent), while others are still uncertain. Finally, ex-post means that a desideratum holds for all possible realizations of the world; thus, it is the strongest notion. However, in a dynamic system, ex-post is often too demanding; but it may still serve as a useful benchmark. 
\end{remark}

\subsection{Economic Efficiency} 
\label{sec:efficiency}
A central goal for the design of a UTM system is to ensure that the airspace is used \textit{efficiently}. Efficient markets maximize value creation for participants and tend to attract the largest number of users. However, this raises the question of how to measure efficiency.

The strongest notion of efficiency is to maximize \emph{welfare}. Different notions of welfare exist, the most common being \emph{utilitarian welfare}, i.e., the sum of all operators' values for the allocation \cite{mas1995microeconomic}.

% The strongest notion of efficiency is to find \emph{efficient} allocations, that is, to ensure that airspace is allocated in such a way that the sum of all operators' values for the allocation, called the \emph{social-welfare}, is maximized.  Unfortunately, this strong notion of efficiency often stands in conflict with other desiderata (citation?) and typically cannot be fully satisfied. Any practical UTM should still aim to minimize the efficiency gap.   
A weaker notion is \emph{Pareto efficiency} \cite{hammond1981ex}. An allocation is Pareto efficient if no operator's flight path can be improved without worsening another operator's flight path. This notion is also well-defined when using mechanisms without money. A further relaxation is \emph{non-wastefulness}, which means that there exist no unallocated resources any agent prefers over (parts of) their allocation.  

The EU's proposed first-come-first-serve mechanism demonstrates how a mechanism may lead to inefficient allocations even if agents report truthfully. To see this, consider an operator who does not have strong preferences regarding the exact route or time of departure (e.g., a surveyor that could fly at any time during the day). If this operator books far in advance, they may block short-term flights with far higher values for their exact route (such as an urgent delivery).
%This might happen even if it is already known that this route is likely to become congested with short-term flights.
The underlying problem is that the mechanism does not make any trade-off decisions. Thus, the first-come-first-serve mechanism does not maximize welfare and might not even be Pareto efficient.
%\footnote{
%or what has been discussed for manned aviation routes and airport slots \cite{ball2006auctions, bichler2021trading}. 

%If operators can reserve routes far in advance, it may be the case that operators booking flights early tend to be flexible, e.g. surveyors, and put lower value on that exact time slot. The more inflexible late bookers, like delivery services, may hold a higher value for flying in that instant, but are inherently disadvantaged in attaining this value. The proposed system would not be maximizing welfare, losing efficiency and fairness.

\subsection{Fairness}
All stakeholders agree that a UTM system should be \emph{fair}. However, there is no single agreed upon notion of fairness \cite{chin2020tradeoffs, evans2020fairness}.

To illustrate the challenge in defining fairness, consider the first-come-first-serve mechanism. Some stakeholders see this mechanism as fair, given that all operators have the same chance to be ``the first.'' Yet, this mechanism disadvantages operators that require shorter planning horizons, which may be considered unfair. Next, consider using an auction that allocates the airspace to the highest bidder. Some see auctions as inherently fair, because no operator gets special treatment and only the bids matter. However, others argue that auctions are unfair given that different operators may have greatly varying financial means. While these disagreements cannot be resolved, there are multiple useful concepts for measuring the fairness of market mechanisms. Ultimately, the regulator must decide which notion to optimize for.

%A very basic notion of fairness, called \emph{anonymity}, requires that the mechanism does not take the operators' identities into account when making an allocation. However, anonymity can collide with the desire for equitable outcomes and is therefore not always desirable. 

One way to measure fairness is via \emph{egalitarian social welfare}, i.e., the minimum %(possibly weighted)
utility of any agent.
%A mechanism that maximizes egalitarian social welfare for all possible reports is also called \emph{max-min fair}.
An alternative notion that puts less weight on single agents is \emph{proportional fairness} \cite{kelly1997charging}. For an allocation to be proportionally fair, there must not exist another allocation for which the
%(possibly weighted) 
sum of each agent's difference in utility is positive. A last notion is \emph{envy-freeness} \cite{foley1966resource}, which means that no agent prefers another agent's outcome. Even though agents have varying preferences, envy-freeness is meaningful as it must hold for \textit{any} possible vector of reports.
%While agents in practice do not typically have values for exactly the same assignment as other agents, this is still a meaningful concept as it has to hold for all possible reports.% As it is a rather strong notion, it might be desirable to relaxe it to almost envy-freeness, such that while agent A could envy agent B, one could eliminate that envy by altering agent B's "better" allocation by a small (bounded) amount. 
%In the context of allocation of discrete bundles, for example by auctioning off blocks of airspace, this would be called \emph{envy-free bounded by a single good} \cite{budish2011combinatorial}. 

\subsection{Simplicity}
All stakeholders we interviewed have emphasized that a \emph{low barrier to entry} is essential to ensure that the airspace is accessible to diverse types of operators. For market design, this implies that the \emph{user interface} to the marketplace must be sufficiently simple \cite{seuken2012market}. 
%Only if the system appears economical to most operators in terms of cost and time investment will it gain the necessary acceptance among regulators and industry.
%Participation in the system should require minimal information and input from the operator.
Standard \emph{direct-revelation} mechanisms \citep{dasgupta1979Implementation} would require operators to report their preferences for all possible flight paths, which is obviously impractical. On the other end of the spectrum are mechanisms (like first-come-first-serve) that ask operators to only state their most preferred flight path, and then either approve or reject their requests. Unfortunately, as explained in Section~\ref{sec:efficiency}, this is inherently inefficient, as it does not allow the mechanism to make any trade-offs. A more principled approach involves designing \emph{smart market mechanisms} \citep{bichler2011SmartMarkets} that provide participants with carefully-designed tools that make it easy for them to (dynamically) report the most important aspects of their preferences \cite{sandholm2013very}, while hiding most of the complexity of the underlying market \cite{seuken2010HiddenMarkets}. Here, AI and machine learning techniques may be used to facilitate autonomous path finding \cite{cieslewski2018data} and to simplify preference reporting \cite{brero2018combinatorial}.

% \subsection{Simplicity (0.4)}
% Stakeholders stressed the need of a low barrier to entry in order to encourage R\&D and participation in the system. Only if the system appears economical to most operators in terms of cost and time investment will it gain the necessary acceptance among regulators and industry. Participation in the system should require minimal information and input from the operator, hiding much of the complexity in order to provide hobbyists access to the market \cite{seuken2010hidden, seuken2012market}. 

% The language in which operators can express their preferences can be chosen to improve user simplicity and computational complexity of the optimisation problem \cite{sandholm2013very}. For instance, one could explicitly rule out preferences that span multiple consecutive flights, or allow preferences over bundles of flights. The right bidding language also helps in finding an optimal allocation and simplifying market clearing. Languages need not be complex to achieve those design goals, as long as operators can fully express their preferences.

\subsection{Incentives}
A key aspect of market design is to ensure that participants have an incentive to report their preferences truthfully. Otherwise, participants may need to spend costly efforts to determine their strategy, and any mechanism that receives manipulated reports will likely make inefficient allocations.

Some of the UTM proposals we have reviewed acknowledge that operators may try to \textit{game} the system \citep{ECUspace}. But instead of disincentivizing such behavior, the proposals foresee a \emph{monitoring and data gathering approach}, with the goal of detecting any manipulative behavior to then correct the mechanisms accordingly. Unfortunately, this is a flawed approach, because by simply analyzing the data, it may be very hard if not impossible to detect what is going wrong. To illustrate this, consider the \emph{school choice problem}, where students are matched with places at competitive high schools \citep{abdulkadiroglu2003SchoolChoice}.
Many cities employ school choice mechanisms that are highly manipulable, such that it is not optimal for students (or their parents) to rank schools in order of preference. Parents have learned how to optimally manipulate the system, ranking a school first that is attractive while also being likely to accept their child. The result is that, when looking at the data, it \textit{seems} as if the vast majority of students receive their first choice, \textit{suggesting} that the mechanism works almost perfectly, even if this is not at all the case \cite{NBERw11965}.
%\cite{dur2018first}.

Market designers are experts in designing mechanisms that provide participants with good incentives for truthful preference reporting. A very strong desideratum is \emph{strategyproofness} \cite{nisan2007stratproof}, which requires that it is optimal for each participant to report truthfully, no matter what the others report. Strategyproof mechanisms are very desirable, as market participants do not need to engage in complex strategizing, which improves the quality of the information the mechanism receives. One can relax this to \emph{(Bayes-Nash) incentive compatibility}, i.e., an agent cannot benefit from misreporting if all other agents report truthfully.

In practice, it is often impossible to guarantee even Bayes-Nash incentive compatibility, in particular in dynamic environments \cite{parkes2010dynamic,Bergeman19}.
%There are also numerous \emph{impossibility results} that show that it is impossible to simultaneously achieve the optimum along multiple design dimensions (like efficiency and strategyproofness), such that trade-offs are always necessary \cite{parkes2001PhDThesis}.
To this end, researchers have proposed various notions of \emph{approximate strategyproofness} \citep{lubin2012approximate, mennle2021partial,azevedo2019strategy}.
%and \emph{strategyproofness in the large} \cite{azevedo2019strategy}, where reporting truthfully becomes approximately ($ex-post$ or $ex-interim$) optimal with a high enough number of competing agents. 
One can also design mechanisms that are computationally hard to manipulate, which provides a certain notion of robustness \cite{faliszewski2010ai}.

Other important considerations are whether a mechanism is \emph{false-name proof} \cite{yokoo2004effect} (i.e., whether agents benefit from ``splitting'' themselves into multiple agents) or \emph{collusion-proof} \cite{ausubel2006lovely}. The latter is important to avoid that incumbent companies coordinate to obtain an advantage over new market entrants. Furthermore, some agents might be \emph{spiteful} \cite{brandt2005spiteful} (i.e., aiming to minimize the utility of competitors). An example of this would be intentionally taking costly detours in order to saturate a certain part of the airspace with the intention of stopping a competitor from flying. Similar behaviour has been observed in manned aviation \cite{valido2020airport}, making this a potential risk in UTM as well. As spiteful behavior may deter competition, which may lead to monopolization and market failure, opportunities for such behaviors should be minimized as much as possible.

% This is an often brought up concern of stakeholders, mostly in the context of monopolistic behaviour. Market participants could be incentivized to act in a way which would not benefit them directly, or even incur a short-term loss, if it meant that they could drive competitors out of the market. For instance, consider a system where conflicts are won by the highest bidder. A delivery company may want to consistently outspend competitors even for flights which they do not have utility for, simply to deny a competitor revenue and causing them to eventually exit the market.

%A strategy-proof system needs to be simple enough for users to understand and accept the strategy-proofness. Recent work on \emph{obvious strategy-proofness} \cite{bade2017gibbardsatterthwaite} attempts to capture the cognitive complexity of a system.

\subsection{Scalability} 
Any UTM system must \textit{scale} well to the ever increasing number of drones.
%This requires a system that allows for the maximum number of flights and does not collapse once the airspace becomes congested.
Scalability as a design goal prevents costly system re-designs in the future. To scale well, the system should be intelligent and largely automated, from soliciting operator preferences to allocating airspace. 

One concern regarding scalability is the \emph{computational complexity} of the employed mechanisms. For example, even if all preferences were known, finding an optimal allocation is typically $\mathcal{NP}$-hard \cite{sandholm2002algorithm, lagoudakis2005auction} and therefore potentially untenable for large, congested markets. One possible direction is to simplify the problem and use heuristic algorithms to find near-optimal solutions to the simpler problem (e.g., \citet{dierksAdmission}). Another option is to aim for coarser, congestion-based allocations and rely on individual drones using autonomous collision avoidance \cite{li2019optimizing}.

\section{Solution Space} \label{solutionspace}

In this section, we provide an overview of possible approaches for designing UTM systems. We divide the solution space into three classes: (1) non-market solutions, (2) markets without money, and (3) markets with money. 

%  A wide range of approaches with varying degrees of centralised control over operators can be envisioned. On one extreme, one can imagine a system without any allocations at all, relying purely on individual UAV autonomously avoiding obstacles and other drones. One could also manage traffic by imposing a topology on the airspace \cite{sunil2015metropolis, joulia2016towards}. Airspace could be sold, for instance by auctioning it off \cite{nakadai} or via congestion pricing as is used in recent proposals for road pricing \cite{cramton2016markets, ostrovsky2019carpooling} or spectrum allocation \cite{cramton2017open}. The system with highest control would be one with centralised navigation, prescribing detailed flight trajectories for each agent \cite{sedov2018centralized}.

\subsection{Non-Market Solutions}
Non-market solutions do not assign airspace based on reports. %This provides an option for the regulator to not concern themselves with the problem of customer facing allocation.
Examples are rule-based systems defining some topology on the airspace akin to road networks \cite{sunil2015metropolis, joulia2016towards, mohamed2018preliminary}, or letting operators autonomously choose their route and avoid collisions. However, since operators have heterogeneous preferences, if the mechanism does not elicit individual preferences, one could end up with a very inefficient allocation or ``traffic jams.'' These approaches are also not able to accommodate well changing numbers of operators or types of routes flown. In an environment that is expected to undergo rapid development in the future, this is not ideal.
Another option is to sell off large parts of the airspace to companies who could then re-sell it, similar to what is done for spectrum licenses \cite{cramton2002spectrum}.
However, note that this gives the resulting intermediaries a lot of power and also tasks them with solving the difficult market design problem.

\subsection{Market-Based Solutions} \label{marketbasedsolutions}

In a market-based system, operators can make reports about their preferences. These might be \emph{direct revelations} of their full preferences, or more limited, for example by only submitting the ideal flight plan. The airspace can then be allocated based on these reports. The more relevant preference information the mechanism has, the better it can optimize the allocation to achieve the market design desiderata described above (e.g., efficiency, fairness, incentives).

%koenig2010progress

\subsubsection{Markets without Money.}
Market mechanisms do not necessarily require monetary transactions. Prominent examples of \textit{markets without money} include (two-sided) matching mechanisms as used in school choice \cite{abdulkadiroglu2003SchoolChoice} as well as (one-sided) assignment mechanisms as used in course allocation \cite{budish2017course}. These mechanisms \emph{do} elicit agents' preferences and then make allocation decisions without charging payments, while still guaranteeing approximate notions of efficiency, fairness, and incentive compatibility. These approaches work well in some settings, for example in schools choice or course allocation, where the students are very homogeneous in their structural needs (e.g., each student is assigned to one school). However, in general, markets without money are relatively limited in their ability to simultaneously achieve multiple desirable properties like efficiency, incentive compatibility and fairness \cite{zhou1990conjecture, yenmez2013incentive, dughmi2010truthful}.

%For example, imagine some operators needing to make thousands of flights per day while others only want to make one flight per month. Here, standard assignment mechanisms that treat all agents the same quickly reach their limits. 

%For instance, any notions of ``urgency'' or ``importance'' (of one flight vs.\ another) cannot easily be expressed in assignment based markets.

% An example of this would be the  first-come-first-serve. While more sophisticated mechanism without money can help avoid equity issues, properly aligning operator incentives in markets without money is hard. Depending on how airspace is allocated, operators may be incentivised to misreport their preferences, claiming to prefer routes that they anticipate to be less popular in order to increase their chances of being assigned. This may seem innocent, but can lead to systematic disadvantages for less strategic agents. A real world example where incentives were neglected is the Boston school choice mechanism, matching students to high schools \cite{abdulkadiroglu2006changing}. 

To handle notions of ``urgency'' or ``importance'' (of one flight vs.\ another), a common approach is to introduce \textit{artificial currency} or \textit{tokens} that operators are assigned by a central entity. These tokens can also be awarded or deducted to reward or punish certain types of behavior within the system, which may help align incentives. A prominent example of such a market is the allocation of food to American food banks \cite{prendergast2017food}. In the context of UTMs, \citet{nakadai} has previously explored this idea. However, one issue that token-based approaches cannot easily solve is that UAV operators may differ greatly in their needs; e.g., some operators may need to make thousands of flights per day while others only want to make one flight per month. To avoid market failure, a large delivery company might therefore require a different number of tokens than a hobby pilot. This raises the challenging question of how many tokens each operator should receive. If one would base the amount of tokens assigned to an operator on the needs the operator \textit{reports}, this would incentivize operators to report higher needs than they actually have. Alternatively, basing token allocation decisions on observable information (e.g., number of flights flown or receipts for drone equipment) would incentivize operators to strategically align their business to inflate this observable characteristic (e.g., flying unnecessary flights). Additionally, any token allocation based on observable information is unlikely to reflect each operator's \textit{actual} needs (e.g., using number of flights flown would disadvantage a wedding photographer with very few, but high value flights).\footnote{To address these problems, one may be tempted to simply increase the amount of observable information taken into account, but this would give administrators the impossible task to anticipate all future use cases and correctly assess their importance and value for society.} While one way to get around this would be to allow operators to \textit{buy} additional tokens with real money, this would effectively turn the system into a market with money.
%\footnote{By controlling and limiting the liquidity of this exchange, regulators could bound how much of an impact different financial budgets can have. However, this approach is limited by the so called \emph{monetary trilemma} \cite{schoenmaker2011financial}.}
%If additional tokens can be bought with real money, an issue arises which in economics is commonly called the \emph{monetary trilemma} \cite{schoenmaker2011financial}. Out of the options of having a fixed exchange rate between money and tokens, introducing a cap on agents' token budgets and letting operators buy an arbitrary amount of additional tokens, a UTM designer with control over the token supply can only ensure two out of the three. 

\subsubsection{Markets with Money.}

Money can help to make trade-off decisions, which can significantly improve efficiency, while keeping the mechanism fair and incentive compatible. For example, if operators need to pay a larger fee the more space they reserve, then they are incentivized not to reserve more space than they need; but if they urgently need a flight, they can decide to pay more. It is worth mentioning that the focus of a UTM system is not to maximize revenue. Payments should therefore only be as high as is necessary to align incentives. In particular, when the airspace is uncongested, payments could be set to zero or some nominal amount.% To this end, the UTM mechanism should probably not charge any payments (or only a nominal fee) if the airspace is uncongested.

Many stakeholders worry that introducing money may present an equity issue, in that players with larger financial value for flights might be able to dominate a market and consistently secure favourable outcomes. However, as has been argued for road pricing, that fear might be overstated \cite{cramton2016markets}. In fact, any revenue that is collected could be used to pay for system costs or could be redistributed to the operators to actually \textit{reduce} inequities between them \cite{levinson2010equity}. It is also important to note that having strong financial means and having a high value for a flight are orthogonal to each other. Large, financially strong players with many flights, such as package delivery services, typically have very small profit margins (on the order of cents), implying a small marginal value per flight. Thus, they do not have an incentive to regularly place large bids to compete with operators such as wedding photographers, who have very high values per flight. Finally, by putting different weights on the requests of different operators, one could improve equity to some degree.

\begin{remark}
For each of the three dimensions \emph{space, time, and degree of centralization}, different levels of granularity seem possible. A mechanism could assign the entire space at once, or subdivide the space into smaller areas to be treated individually. Similarly, airspace could be allocated for long time periods at a time, or on a minute-by-minute basis. Finally, mechanisms could range from a centralized allocation mechanism, prescribing routes and deviations, to drones negotiating only bilaterally. Centralized solutions make designing for efficiency, fairness and good incentives much easier, but a distributed approach may be necessary to manage computational complexity and allow the system to scale. Research in distributed mechanism design exists, but is scarce \cite{nisan2007distributed}.
\end{remark}

\section{Conclusion}
In this paper, we have introduced drone traffic management as a market design problem of high practical importance. We have outlined why it is essential for the UTM community (and in particular for regulators) to call on the expertise of AI and market design researchers to ensure that future UTM systems are well designed. For researchers, the next years present a prime opportunity to have a practical impact, as many countries are currently establishing or overhauling their UTM systems. One challenge when trying to influence policy will be to convince policy makers that any new solution offers significant advantages in practice. It will therefore not suffice to only propose and study better mechanisms, but a clear focus also needs to be placed on communicating the advantages and disadvantages of different approaches to audiences outside the research community.

\section*{Acknowledgments} \label{acknowledgements}
We are thankful to Larissa Haas and Benoit Curdy from the Swiss Office of Federal Aviation for introducing us to the UTM problem, for putting us in touch with various stakeholders, and for the pleasant collaboration. Furthermore, we thank (in alphabetical order) Christoph Derrer (Swiss Post), Antony Evans (Airbus UTM), Mike Glasgow (Wing), Andrew Hately (EUROCONTROL), Ralf Heidger (DFS), Pierre-Alain Marchand (senseFly), Shinji Nakadai (NEC Corp.), Davide Scaramuzza (UZH Robotics and Perception Group), and many others for useful discussions. All views expressed in this paper are our own and do not necessarily represent the views of the people we interviewed. Part of this research was supported by the European Research Council (ERC) under the European Union's Horizon 2020 research and innovation programme (Grant agreement No. 805542).

\newpage
% Use \bibliography{yourbibfile} instead or the References section will not appear in your paper
\bibliography{aaai22}

\end{document}